\documentclass[aps,prl,reprint,nobibnotes, longbibliography]{revtex4-1}

\usepackage{t1enc}
\usepackage[utf8]{inputenc}
\usepackage{amsmath}
\usepackage{graphicx}
\usepackage{enumerate}
\usepackage{natbib}
\usepackage{multibib}
\newcites{M}{references}
\newcites{S}{references_supplement}
\begin{document}

\title{1 minute parity lifetime of a NbTiN Cooper-pair transistor}

\author{David J.~van Woerkom}
\thanks{These authors contributed equally to this work.}
\author{Attila Geresdi$^*$}
\email[Corresponding author, e-mail address: ]{a.geresdi@tudelft.nl}

\author{Leo P.~Kouwenhoven}
\affiliation{QuTech and Kavli Institute of Nanoscience, Delft University of
Technology, 2600 GA Delft, The Netherlands.}

\maketitle

\textbf{The parity modulation of the ground state of a superconducting island is
a direct consequence of the presence of the Cooper pair condensate preferring an
even number of charge carriers \citeM{PhysRevLett.69.1993, Lafarge1993}. The
addition energy of an odd, unpaired quasiparticle equals to the superconducting
gap, $\Delta$, suppressing single electron hopping in
the low temperature limit, \(k_B T \ll \Delta\). Controlling the quasiparticle
occupation is of fundamental importance for superconducting qubits as single
electron tunneling results in decoherence \citeM{Clarke2008, Riste2013}.
In particular, topological quantum computation relies on the parity control and
readout of Majorana bound states \citeM{0034-4885-75-7-076501, read2012}.
Here we present parity modulation for the first time of a niobium titanite
nitride (NbTiN) Cooper-pair transistor coupled to aluminium (Al) leads. We
show that this circuit is compatible with the magnetic field requirement \(B
\sim 100\,\)mT of inducing topological superconductivity in spin-orbit coupled
nanowires \citeM{PhysRevLett.105.077001, PhysRevLett.105.177002, Mourik25052012}.
Our observed parity lifetime exceeding 1 minute is several orders of magnitude
higher than the required gate time of flux-controlled braiding of Majorana
states \citeM{PhysRevB.88.035121}. Our findings readily demonstrate that a NbTiN
island can be parity-controlled and therefore provides a good platform for
superconducting coherent circuits operating in a magnetic field.}

Experimentally, the parity modulation of a superconducting island can be
observed via the ground state charge \citeM{Lafarge1993}, the even-odd modulation
of the charge stability diagram \citeM{PhysRevLett.69.1997, PhysRevLett.70.1862},
or the parity dependence of the switching current, $I_\textrm{sw}$
\citeM{PhysRevLett.72.2458}. The interplay
of the charging energy \(E_c = e^2/2C\) and the Josephson coupling \(E_J=I_c
\hbar /2e\) makes the Cooper-pair transistor (CPT) a single,
gate-modulated Josephson junction \citeM{PhysRevLett.65.377,
PhysRevLett.72.2458} with a \emph{2e} charge periodicity in the absence of
parity switches, i.e.~infinitely long parity lifetime, \(\tau_p\).

\begin{figure}[ht!]
\centering
\includegraphics[width=0.5\textwidth]{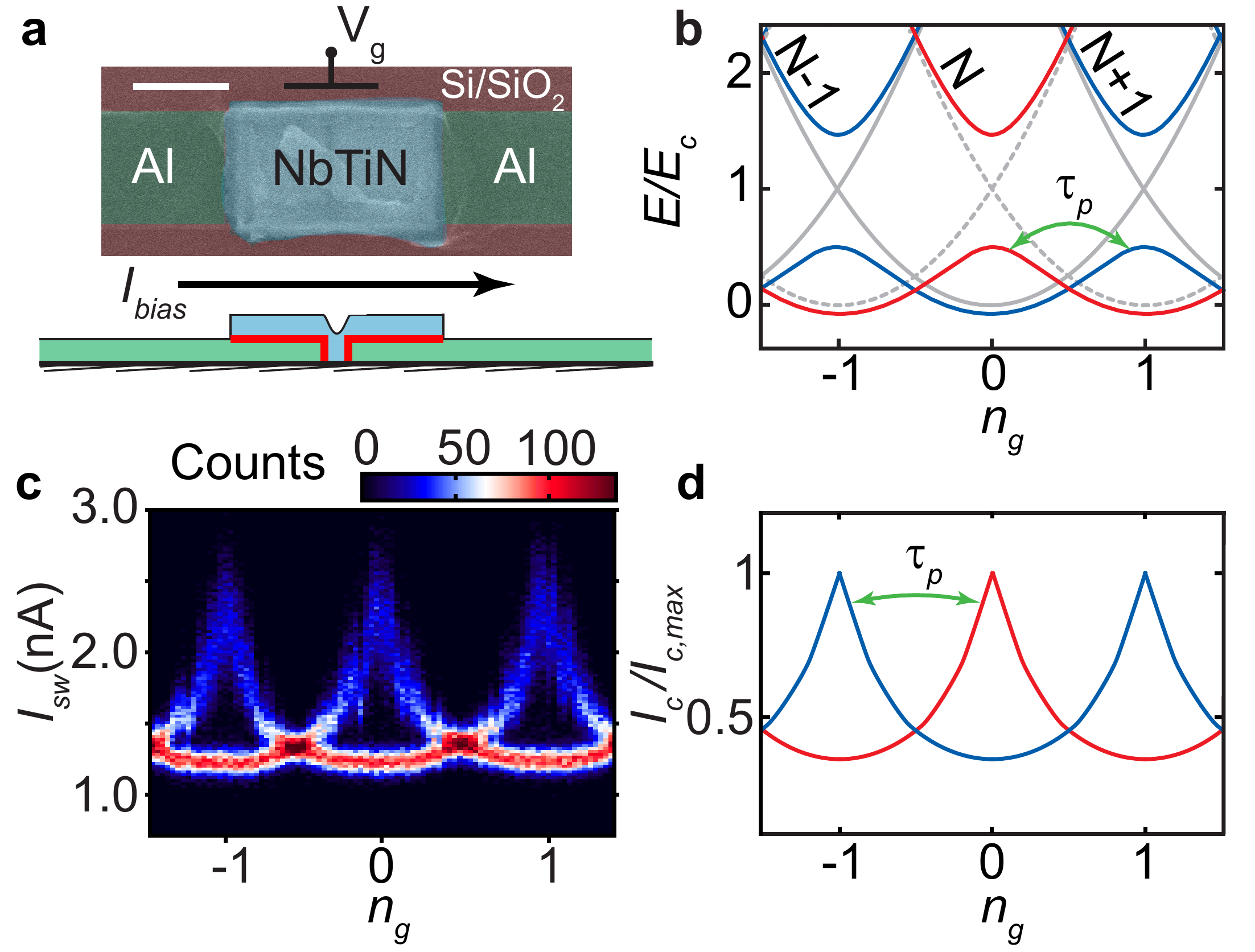} \caption{\textbf{Device
layout and parity-dependent switching current.} \textbf{(a)} Scanning
electron microscope image and schematic cross-section of a typical Al/NbTiN/Al
hybrid Cooper-pair transistor (CPT). The NbTiN island dimensions are
$250\,$nm$\,\times\,450\,$nm.
Scale bar denotes $200\,$nm.
The AlO$_x$ oxide barriers are indicated by thick red lines. \textbf{(b)} Energy
level diagram as a function of the gate charge, $n_g$, in the presence of low energy subgap
states restoring \emph{1e} periodicity.
Gray lines denote energy levels in the absence of Josephson coupling,
i.e.~\(E_J=0\). Red and blue lines show energy levels for even and odd charge
parity respectively, both for \(E_J=E_c\). Parity switches occur on the
timescale of \(\tau_p\) due to quasiparticle tunneling. Measured switching current histogram
\textbf{(c)} and calculated $I_c(n_g)$ \textbf{(d)} in the low temperature
limit. Note that in (d) the two possible $I_c(n_g)$ values corresponding to the
even and odd charge states denoted by blue and red lines respectively. In the
measured data (c) the two branches are superimposed, see text.}
\end{figure}

\begin{figure*}
\centering
\includegraphics[width=\textwidth]{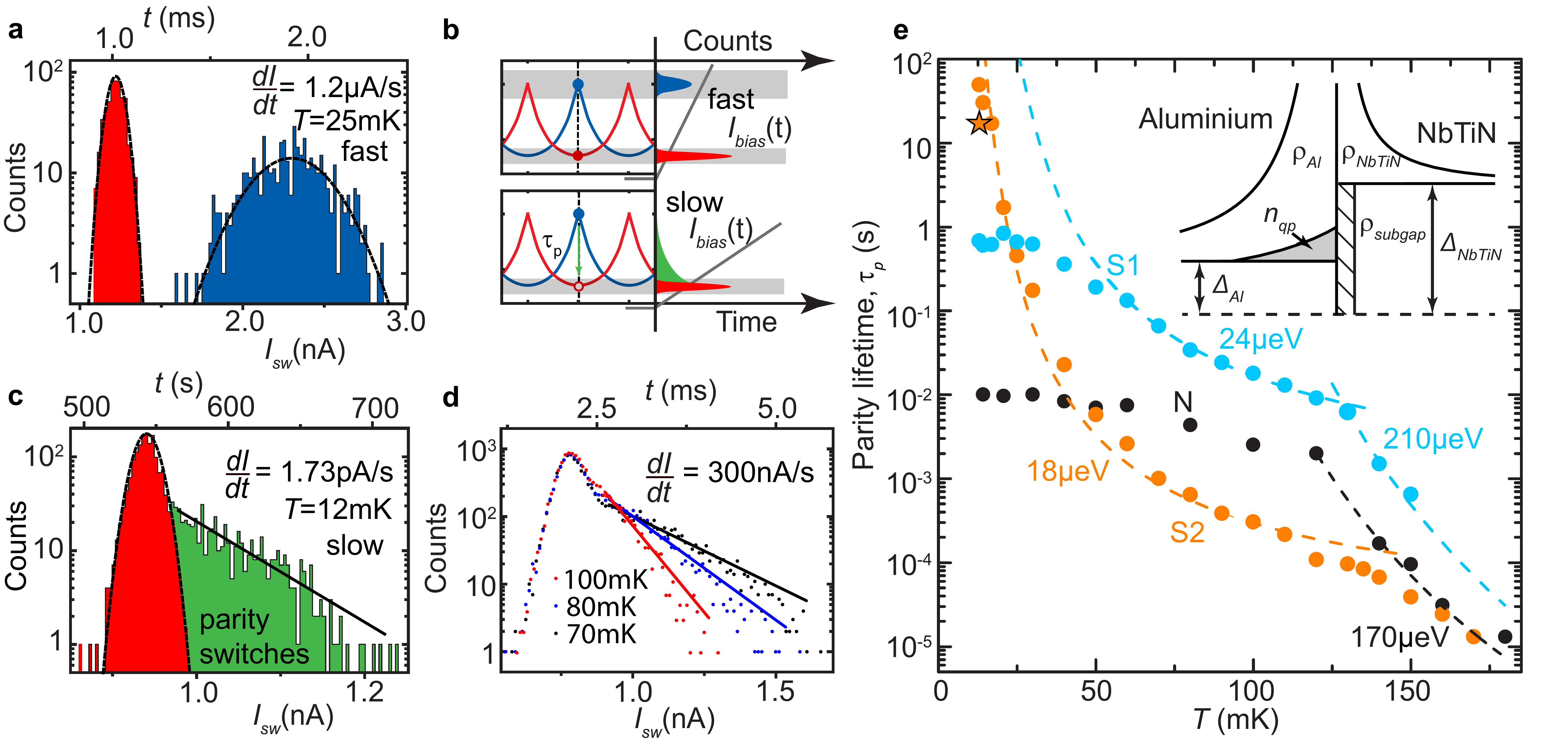}
\caption{\textbf{Characterization and temperature dependence of the parity
lifetime.} \textbf{(a)} Linecut histogram at integer $n_g$ of Fig.~1c showing a
bimodal distribution. We attribute the two peaks to the two parity states of the
CPT (colored red and blue, respectively). \textbf{(b)} For a \emph{fast}
current ramp (upper panel), the histograms of the two parity states are
independently probed showing the characteristics of data in panel (a). In the
\emph{slow} limit (lower panel in (b)), parity switches occur during the current
ramp, leading to an exponential tail of the distribution (shown in green),
quantifying $\tau_p$. The $I_\textrm{bias}(t)$ current ramp is represented by a
dark gray line.
\textbf{(c)} Experimental dataset in the \emph{slow} limit.
Note the change in the current- and timescale compared to panel (a). We show the
exponential cutoff in green, and extract $\tau_p=49\,$s (solid black line).
\textbf{(d)} Experimental data at different temperatures show the temperature
dependence of $\tau_p$. \textbf{(e)} $\tau_p$ as a function of temperature for
non-shielded device N, and shielded devices S1, S2. For detailed comparison, see
the main text and the supplementary material.
All CPTs exhibit an activated behavior with \(\Delta\approx 170\ldots 210\mu\)V
in the high temperature limit corresponding to the gap of the aluminium leads.
Saturation of device N without shielding and no quasiparticle traps is observed
below \(T \approx 100\,\)mK. Shielded devices S1 and S2 exhibit a
minigap-activated behavior \(\Delta^\star \approx 20\,\mu\)V in the low
temperature limit. The fitted $\tau_p(T)$ function is defined in the main text.
Star symbol shows \(\tau_p=15\,\)s at \(T=12\,\)mK extracted from parity
distilled data for device S2 (see Fig.~3d).}
\end{figure*}

Recent, direct measurements of \(\tau_p\) \citeM{PhysRevB.78.024503, Riste2013}
yielded values up to the millisecond regime for aluminium devices. Despite
considerable efforts, however, no \emph{2e} periodicity has been reported for
non-aluminium superconductors \citeM{Dolata2005, Savin2007, PhysRevB.76.172505},
such as niobium or vanadium. Comparative studies of aluminium and niobium
CPTs suggested that the elusiveness of parity effects is related to the material
properties \citeM{Savin2007}, in accordance with earlier measurements showing
that subgap quasiparticle states may appear in niobium due to oxidization of the bulk
material \citeM{Halbritter1987}. In contrast, nitridized niobium compounds, such
as niobium titanite nitride (NbTiN), have been shown to be less prone to form
oxides and hence are good candidates for parity-conserving superconducting
circuits. Furthermore, NbTiN forms transparent contacts with
spin-orbit coupled semiconductor nanowires \citeM{Mourik25052012}, and has become
a preferred superconductor to investigate Majorana bound states.

Our device features a NbTiN island sputtered onto Al leads (Fig.~1a). The tunnel
barriers between the island and the leads are created by means of controlled
\emph{in-situ} surface oxidization of Al, resulting in amorphous AlO$_x$
barriers \citeM{Oliver2013}.

We extract a charging energy \(E_c\approx50\,\mu\)eV from the
measured charge stability diagram.
For different devices, we estimate \(E_J\approx 30\ldots 50\, \mu\)eV from the
superconducting gaps and normal state resistances of the junctions
\citeM{PhysRevLett.10.486} assuming equal resistances for the two tunnel
barriers. A detailed list of parameters and characterization methods are
presented in the supplementary material.
Our devices are in the intermediate coupling regime with \(E_J
\sim E_c\), where the energy diagram (Fig.~1b) and the critical current
(Fig.~1d) are sensitive to the charge parity. It is important to
note that our CPTs are in the optimal regime to establish flux-controlled
braiding of Majorana bound states with \(E_J \sim E_c \gg k_B T\)
\citeM{PhysRevB.88.035121} and hence a useful platform to establish the
parity lifetime for Majorana circuits \citeM{PhysRevB.85.174533}.

We model the CPT as a two level system which can exist in either parity state
(red and blue bands in Fig.~1, respectively), and switches state on the
timescale of $\tau_p$ \citeM{PhysRevB.74.064515} due to quasiparticle tunneling.
We collect the switching current histograms by repetitively sweeping the bias
current from zero (non-dissipative state) to beyond the switching current. Here,
in the resistive state, quasiparticle tunneling causes a random reinitialization
of the parity state of the CPT for the next measurement. This results in the
apparent \emph{1e} periodicity in Fig.~1c. Nevertheless, as long as the parity
remains constant during each sweep, we expect to find the two branches as a
bimodal histogram, as we indeed observe in Fig.~2a. In these measurements, the
current ramp time is much shorter than the parity lifetime, $\tau_p$
(\emph{fast} measurement limit).

We quantify \(\tau_p\) in the \emph{slow} measurement limit. In this regime
parity switches occur during the current ramp (Fig.~2b lower panels) such that
reaching the upper branch (depicted as blue in
Fig.~2a and 2b) becomes exponentially suppressed (Fig.~2c). The
exponential tail represents parity switches during the current bias ramp,
resulting in an observable decay of the upper branch (depicted as green in
Fig.~2b and 2c), \(p_u(t)=p_u(0)\exp(-t/\tau_p)\). Thus, from the decay of the
histogram (black solid line in Fig.~2c), we can directly obtain $\tau_p$.

The observed \(\tau_p\) is a result of single electron tunneling events through
the junctions of the CPT. In the zero voltage state, we can link \(\tau_p\) to
the subgap density of states on the island and the quasiparticle density in the
leads \citeM{PhysRevB.85.012504} (for details see the supplementary material):
\begin{equation}
\tau_p^{-1}=\frac{2n_\textrm{qp}}{e^2R_N\rho_\textrm{Al}}\frac{\rho_\textrm{subgap}}{\rho_\textrm{NbTiN}},
\end{equation}
where $R_N$ is the normal state resistance of the CPT. It is instructive to note
that the parity lifetime is determined by the quasiparticle density in
the leads \citeM{PhysRevB.85.174533} and the \emph{phenomenological} Dynes
parameter \citeM{PhysRevLett.41.1509} of the island material. Assuming a thermal
\(n_\textrm{qp}(T) \propto \sqrt{T}\exp(-\Delta_\textrm{Al}/k_B T)\) in the
leads, we find \(\Delta_\textrm{Al}=170\ldots 210\,\mu\)eV (Fig.~2e) for temperatures exceeding
$120\,$mK, in good agreement with the superconducting gap of the aluminium leads
extracted from the charge stability diagram. We can therefore attribute the
observed parity lifetime for $T>120\,$mK to the \emph{thermal} quasiparticle
population in the leads.

For device N, however, we find a saturated $\tau_p=9.5\,$ms in the low
temperature limit, a common observation in superconducting qubits
\citeM{Riste2013} and hybrid single electron transistors
\citeM{PhysRevB.85.012504} signifying the presence of \emph{non-thermal}
quasiparticle excitations.

We further improve the low temperature parity lifetime by introducing
microwave-tight shielding coated with infrared absorber painting and Ti/Au
quasiparticle traps for devices S1 and S2. These additions result in a
non-saturated behaviour of $\tau_p$, and we observe a minigap activated
behaviour with \(\Delta^{\star}\approx 20\,\mu\)eV for both devices. It is to be
stressed that this observation signals that the effective quasiparticle
temperature of the CPT follows the bath temperature down to the $10\,$mK regime.
We find \(\tau_p=49\,\)s at \(T=12\,\)mK for device S2. To put this
number into context, we note that the Josephson frequency $f_J=E_J/h \approx
10\,$GHz and thus a single quasiparticle event occurs only once for every
\(\tau_p f_J \sim 10^{11}\) Cooper pairs tunneling through the junctions. This
signifies the low probability of parity switches for an open device with \(E_J \approx
E_c\) required for flux-tunable Majorana braiding schemes.

\begin{figure}[ht!]
\centering
\includegraphics[width=0.5\textwidth]{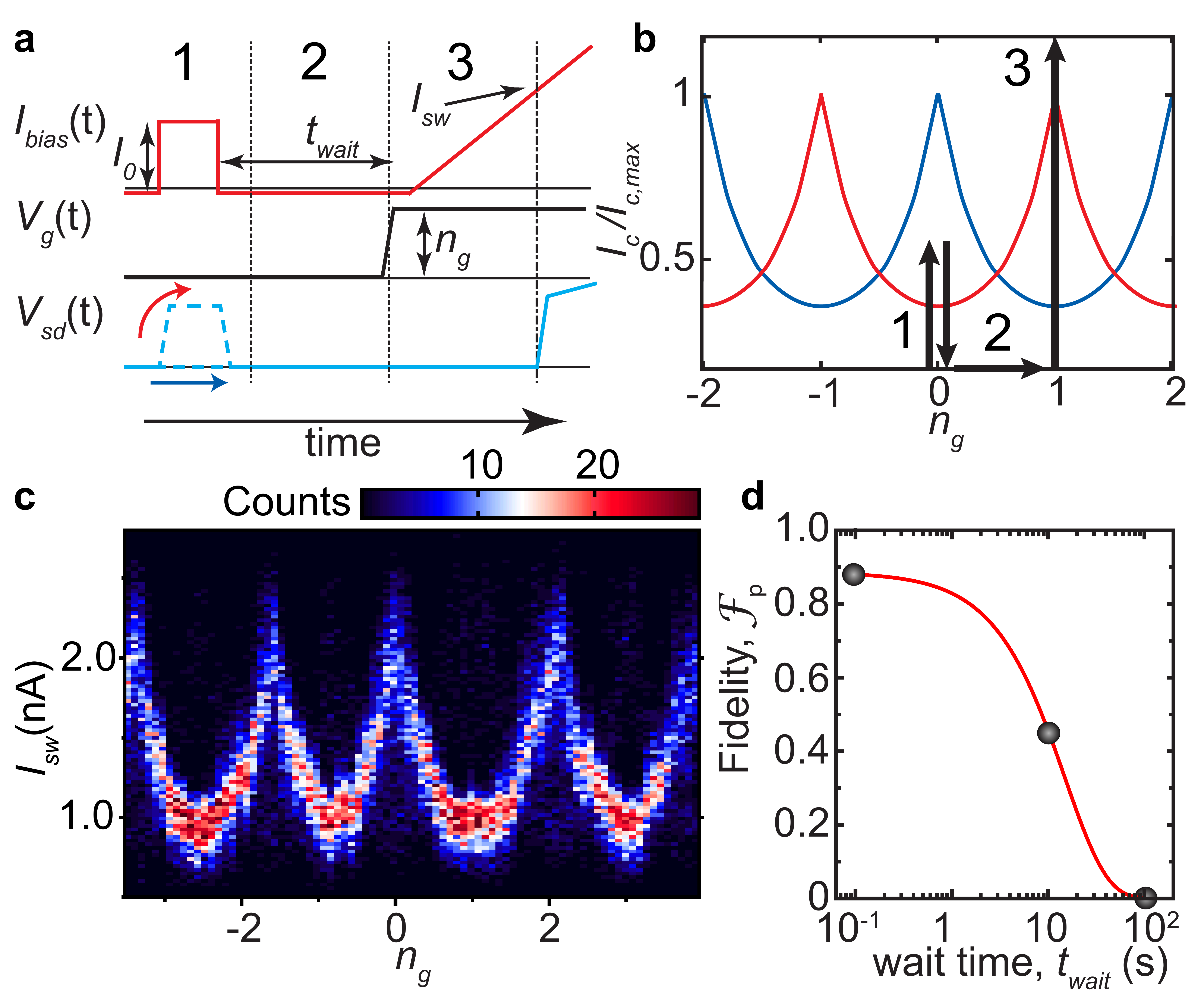}
\caption{\textbf{Parity state distillation.} \textbf{(a)} Schematic current bias and gate
voltage waveforms applied to the device. First, current
bias pulse with an amplitude between the two branches is applied (1). At the
same time the measured voltage \(V_\textrm{sd}(t)\) is recorded to initialize the parity state. Then the gate
voltage is ramped (2) and finally, the switching current is recorded (3). \textbf{(b)} The
schematic representation of the applied waveforms overlayed on the
parity-dependent critical current.
\textbf{(c)} Parity distilled switching current histogram exhibiting \emph{2e}
periodicity with \(t_\textrm{wait}=100\,\)ms. \textbf{(d)} Parity distillation fidelity as
a function of \(t_\textrm{wait}\), see main text. Solid red line
denotes the fit \(\sim \exp(-t_\textrm{wait}/\tau_p)\) with \(\tau_p=15\,\)s.
All data was recorded on device S2.}
\end{figure}

Thus far, we started each switching current measurement from an unknown parity
state because of the random reinitialization in the dissipative state of the
CPT for $I_\textrm{bias}>I_\textrm{sw}$. In order to reproducibly select the same
parity state, we employ a parity distillation protocol (Fig.~3) where, by
selecting a single parity state without switching to the resistive state, we ensure
that the parity remains well defined for the subsequent measurement. This
protocol indeed results in a \emph{2e} periodic switching current pattern
(Fig.~3c) which is observed for the first time for a non-aluminium CPT.

We quantify the effectiveness of the parity distillation by defining the
fidelity as:
\begin{equation}
\mathcal{F}_p=\frac{p_{u,f}-p_{u,i}}{1-p_{u,i}},
\end{equation}
where $p_{u,f}$ is the conditional probability of the upper branch in the final
step (3), and $p_{u,i}$ is the initial probability. We note that the above
expression is valid for an arbitrary $0<p_{u,i}<1$ value set by the average
quasiparticle occupation of the CPT. For device S2 we find
\(\mathcal{F}_p=0.88\pm0.05\) for \(t_\textrm{wait}=100\,\)ms demonstrating
the high degree of parity distillation. By changing $t_\textrm{wait}$ between
the parity initialization (1) and measurement (3), we acquire an independent measurement
\(\tau_p=15\,\)s for device S2 at $T=12\,$mK (Fig.~3d).

\begin{figure}[ht!]
\centering
\includegraphics[width=0.5\textwidth]{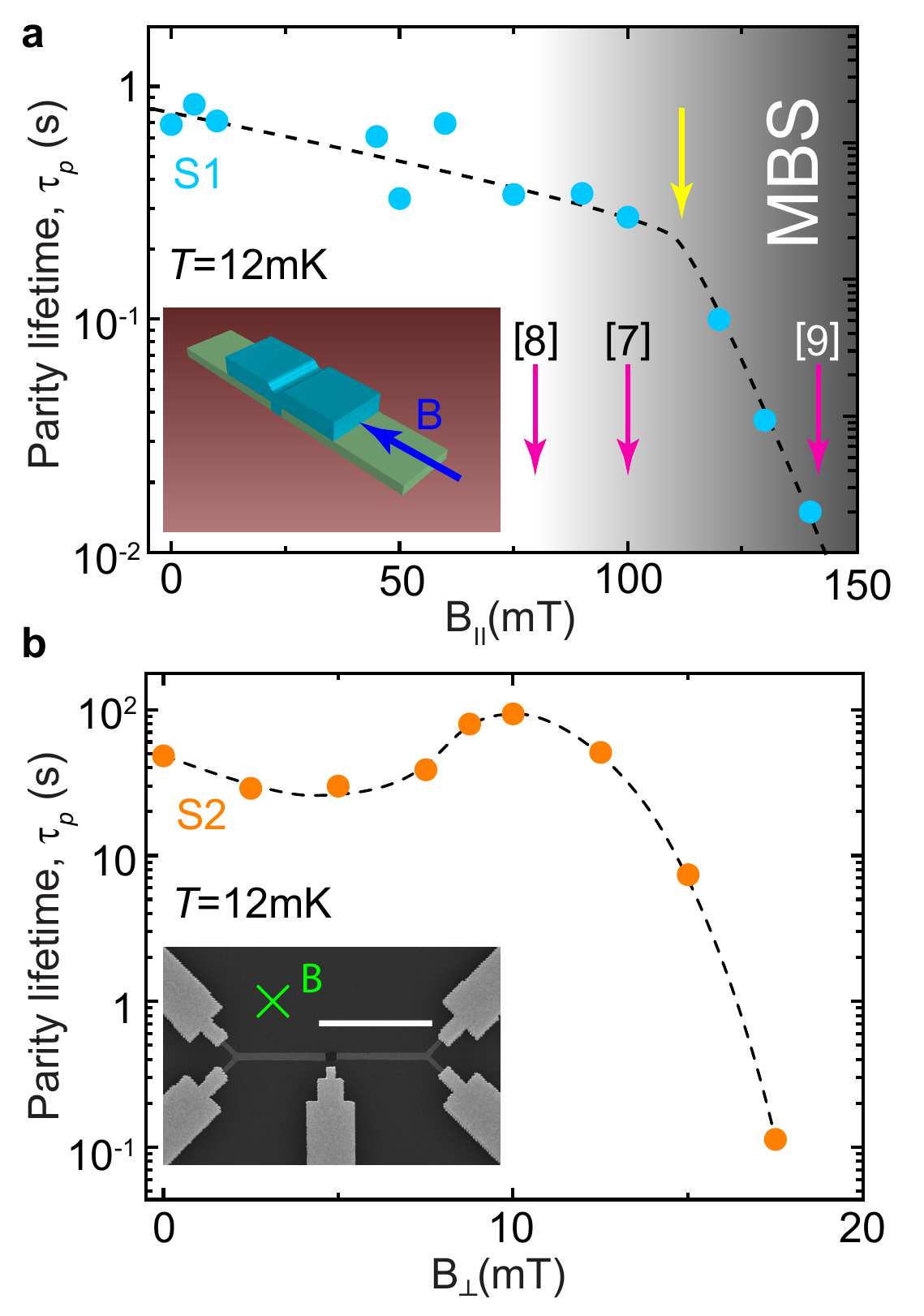}
\caption{\textbf{Influence of the magnetic field on the CPT.} \textbf{(a)}
The parity lifetime as a function of the in-plane field, $B_\parallel$ pointing
as shown in the inset. The shaded region depicts the condition for inducing Majorana bound
states (MBS) in spin-orbit coupled semiconductor
wires \protect\citeM{PhysRevLett.105.077001, PhysRevLett.105.177002,
Mourik25052012}.
The change in slope at $B\approx110\,$mT is denoted by the yellow arrow and the dash
line serves as a guide to the eye. \textbf{(b)} The parity lifetime as a
function of perpendicular field, $B_\perp$. The dash line shows the overall
trend. The inset shows the electron microscope image of the device with the white scale bar denoting $5\,\mu$m.}
\end{figure}

Finally, we investigate the evolution of \(\tau_p(B)\) in different
magnetic field directions. In parallel magnetic field, we observe a gradual
decrease of \(\tau_p\). The onset of the steep decay at $B=110\,$mT (yellow arrow in Fig.~4a) is
in agreement with the condition for vortex penetration through a
mesoscopic superconducting island \citeM{Moshchalkov1995} with \(\Phi\approx
1.1\Phi_0 \gtrsim\Phi_0\). This result underlines the significance of the sample
geometry for magnetic field-enabled CPTs to avoid vortex formation. However, our
device exhibits \(\tau_p>10\,\)ms in \(B_\parallel>100\, \)mT, 
required to induce Majorana bound states \citeM{PhysRevLett.105.077001,
PhysRevLett.105.177002, Mourik25052012}.

In contrast, applying a perpendicular field first results in an increase
of \(\tau_p\) reaching a maximum value \(\tau_p=94\,\)s at \(B_\perp\approx
10\,\)mT, before dropping at higher magnetic fields (Fig.~4b). Making use of the
relation between the lower critical field and the stripe width, \(B_{c1}(w)\sim
\Phi_0/w^2\) \citeM{PhysRevLett.92.097003} we attribute the initial increase to
vortex formation and hence more effective quasiparticle trapping in the wide
lead sections (\(w\approx 2\,\mu\)m) of the device (see wide light gray sections in Fig.~4b).
Upon reaching \(B \approx 10\,\)mT, the vortex phase becomes stable in the close
vicinity of the CPT (\(w\approx 250\,\)nm) causing a gradual decrease of
\(\tau_p\). We note that due to \(\xi_{Al}\approx 100\,\textrm{nm} \sim w\) for
our devices, we cannot make a quantitative comparison between our measurement
data and theoretical expressions of critical magnetic field of thin stripes.
We however conclude that vortices induced by a perpendicular magnetic field can
increase the efficiency of quasiparticle traps, but the formation of a vortex
phase in the near vicinity of the CPT enhances quasiparticle transport in
agreement with earlier observations \citeM{PhysRevB.84.220502, wang2014}.

In conclusion, we fabricated and characterized NbTiN Cooper-pair transistors
showing parity effects for a non-aluminium superconductor for the first time.
We characterize the parity lifetime by evaluating the switching current
histograms and find values exceeding 1 minute at \(T=12\,\)mK. Our devices
are in the regime of \(E_j \sim E_c \gg k_B T\) ideal for Majorana braiding schemes.
Furthermore, we demonstrate charge parity distillation to reproducibly
initialize the island in a given parity state. We find parity lifetimes in
excess of $10\,$ms in external magnetic fields up to $150\,$mT and showed the
importance of sample geometry for magnetic field-enabled operation 
required for inducing Majorana bound states.

The authors thank A.~R.~Akhmerov, S.~Rubbert, Y.~Nazarov, R.~Lutchyn and
J.~Pekola for fruitful discussions and R.~N.~Schouten for technical assistance. This work
has been supported by the Netherlands Foundation for Fundamental
Research on Matter (FOM) and Microsoft Corporation Station Q. A.~G.~acknowledges
funding from the Netherlands Organisation for Scientific Research (NWO) through
a VENI grant.

D.~J.~W.~fabricated the devices. D.~J.~W.~and A.~G.~performed the measurements.
D.~J.~W., A.~G.~and L.~P.~K.~discussed the data, contributed to the analysis and
wrote the manuscript.

\bibliographyM{references}

\onecolumngrid
\pagebreak

\begin{center}
\textbf{\large Supplementary online material} 

\textbf{\large 1 minute parity lifetime of a
NbTiN Cooper-pair box}
\end{center}

\setcounter{figure}{0}
\makeatletter 
\renewcommand{\thefigure}{S\@arabic\c@figure}
\makeatother

\setcounter{equation}{0}
\makeatletter 
\renewcommand{\theequation}{S\@arabic\c@equation}
\makeatother

\setcounter{table}{0}
\makeatletter 
\renewcommand{\thetable}{S\@arabic\c@table}
\makeatother

\setcounter{page}{1}

\section{Device fabrication}

The Cooper-pair transistors (CPTs) were fabricated using electron beam
lithography and thin film deposition starting with a p$^{++}$ doped silicon
wafer with a $285\,$nm thick thermally grown SiO$_2$ surface layer. First,
aluminium leads were defined and evaporated in a high-vacuum chamber
($p_\textrm{base}\sim10^{-7}\,$Torr) at a rate of $0.2\,$nm/s with a thickness
of \(30\ldots35\,\)nm. Subsequently, the mask for the NbTiN island was defined
in a second lithography step. The sample was loaded into an ultra high vacuum
AJA International ATC 1800 sputtering system
($p_\textrm{base}\sim10^{-9}\,$Torr), where first a $\sim 5\,$nm Al layer was
removed by means of argon plasma etching at $p=3\,$mTorr. This step was followed
by \emph{in-situ} oxidization to create AlO$_x$ tunnel barriers.
Without breaking vacuum, the NbTiN island was then sputtered with a layer
thickness of $70\ldots 100\,$nm.

We used a Nb$_{0.7}$Ti$_{0.3}$ target with a diameter of $3''$.
Reactive sputtering resulting in nitridized NbTiN films was performed in an Ar/N
process gas with a typical $10\,at\%$ nitrogen content at a pressure of
$3\,$mTorr using a DC magnetron source. During deposition, a $-45\,$V bias was
applied to the sample with respect to the target. The critical temperature of
the superconducting transition temperature of thin films with a layer thickness
of $100\,$nm was measured to be $14.1\,$K in zero magnetic field.

For the shielded samples S1 and S2, quasiparticle traps were fabricated by first
cleaning the Al surface by means of argon plasma milling at $p=0.2\,$mTorr and
\emph{in-situ} evaporation of $25\,$nm Ti and $100\,$nm Au films at a base
pressure of $\sim10^{-7}\,$Torr (see Fig.~S1c and d).

Care was taken to remove resist mask residues after each electron beam writing
step using a remote oxygen plasma etch with a pressure of $1\,$mbar.

\begin{table*}[!ht]
\centering
\begin{tabular}{| c || c | c | c | c | c | c |}
\hline
device & island size & junction size  & NbTiN thickness & traps & oxygen
exposure & $R_N$ \\
& (nm$\,\times\,$nm) & (nm$\,\times\,$nm) & (nm) & & (Torr$\times$s) &
(k$\Omega$)\\
\hline
\hline
N & $500\,\times\, 500$ & $200\,\times\, 200$ & 70 & no & 7400 & 58 \\
S1 & $450\,\times\, 200$ & $200\,\times\, 200$ & 100 & yes & 150 & 125\\
S2 & $450\,\times\, 250$ & $200\,\times\, 250$ & 100 & yes & 150 & 66 \\
\hline
\end{tabular}
\caption{\textbf{Device fabrication parameters of the CPTs discussed in
the main text.}}
\end{table*}

\begin{figure}[ht!]
\centering
\includegraphics[width=\textwidth]{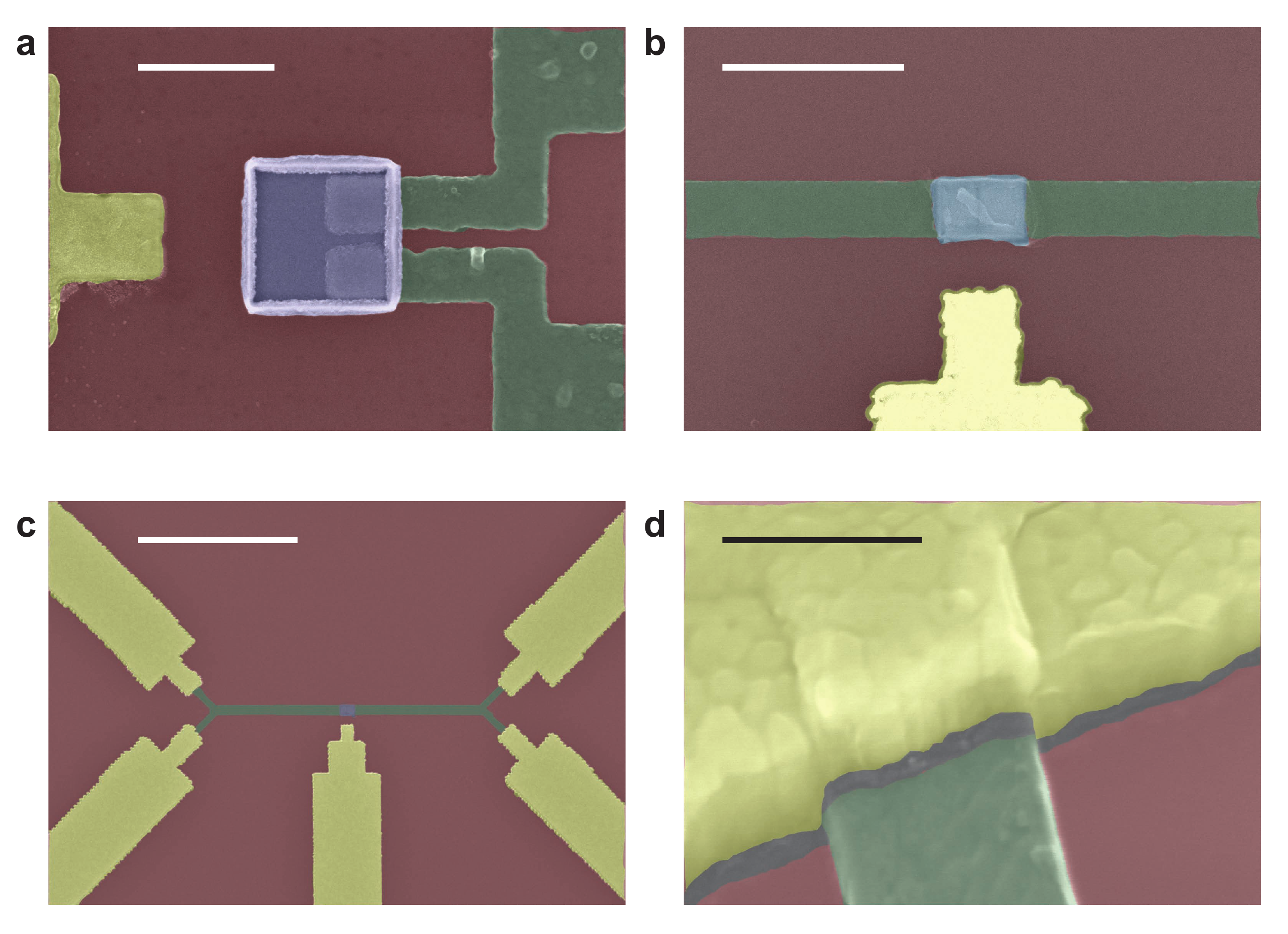}
\caption{\textbf{Scanning electron microscope images of the CPTs.}
\textbf{(a)} Non-shielded device N without quasiparticle traps. \textbf{(b),
(c), (d)} Shielded device S2 featuring quasiparticle traps.
Colour legend: dark red:
Si/SiO$_2$ substrate; light blue:
NbTiN island, green: aluminium leads and yellow:
normal metal (Ti/Au) gate and quasiparticle traps. Scale bars denote $500\,$nm
\textbf{(a)}, $1000\,$nm \textbf{(b)}, $5\,\mu$m \textbf{(c)} and $200\,$nm
\textbf{(d)}, respectively.}
\end{figure}

\section{Measurement setup}

The measurements were performed in a Leiden Cryogenics CF-1200 dry dilution
refrigerator with a base temperature of $12\,$mK equipped with Cu/Ni shielded
twisted pair cables thermally anchored at all stages of the refrigerator.

\begin{figure}[ht!]
\centering
\includegraphics[width=0.7\textwidth]{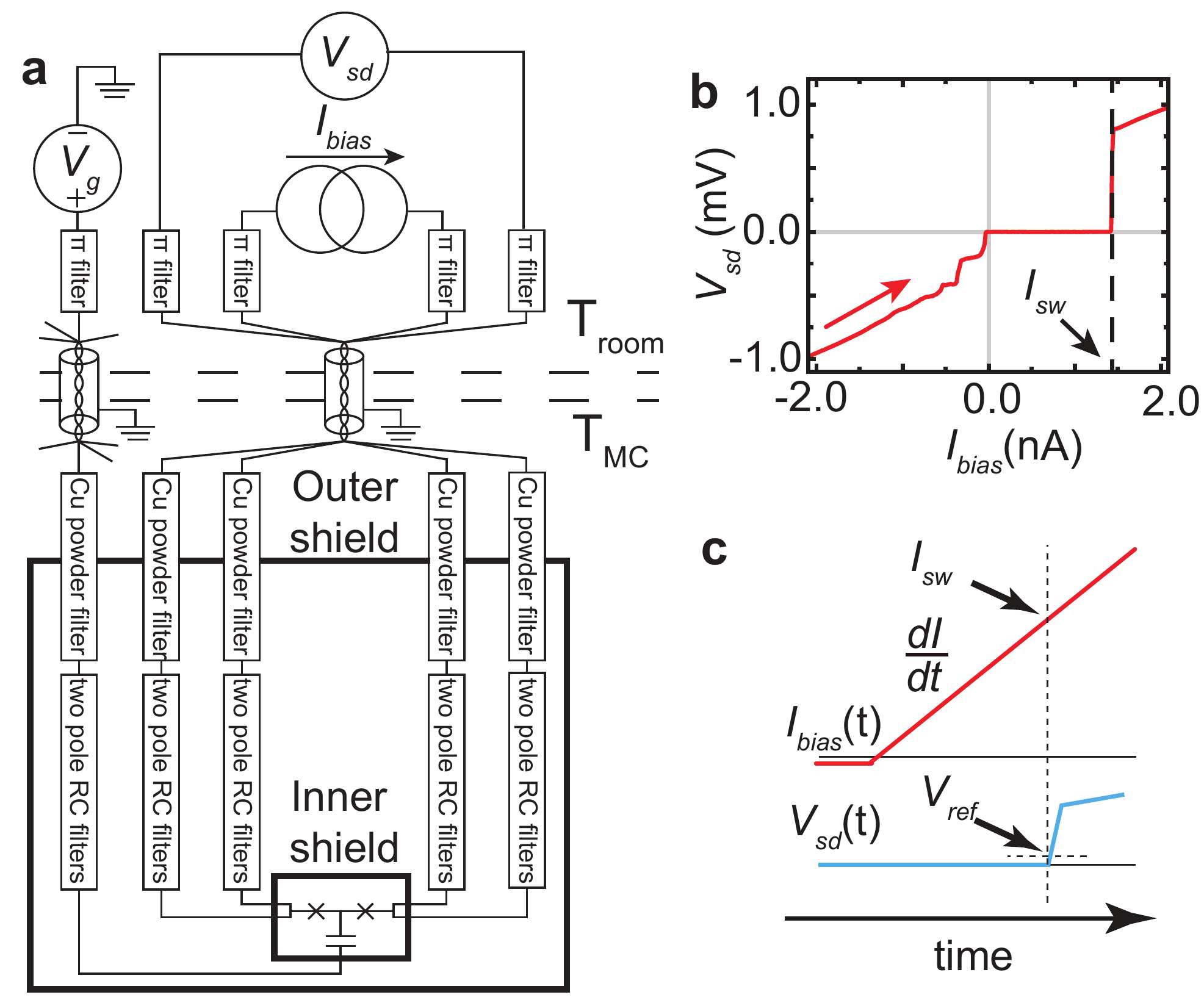}
\caption{\textbf{Measurement electronic setup and typical waveforms.}
\textbf{(a)} Schematics of the measurement. \textbf{(b)} Typical
experimental $V-I$ trace exhibiting a well-defined switching current,
$I_\textrm{sw}$. \textbf{(c)} Current ramp (red) with a slope of $dI/dt$ probing
$I_\textrm{sw}$ which is recorded when the measured $V_\textrm{sd}$
(blue) reaches $V_{ref}\sim 10\,\mu$V.}
\end{figure}

Fig.~S2a shows the schematics of the measurement. The current bias and gate
voltage were applied through battery operated and optically isolated sources in
order to reduce line interference. Similarly, the first stage of the
$V_\textrm{sd}$ amplifier was isolated from the commercial readout electronics.
Filtering of the measurement lines was achieved by room temperature LC $\pi$
filters with a cutoff frequency of $\sim 100\,$MHz followed by a sequence of a 
high frequency copper powder filter
\citeS{s:/content/aip/journal/rsi/84/4/10.1063/1.4802875} and a two-pole RC
filter with a nominal $f_{-3dB}=50\,$kHz, both thermally anchored to the $12\,$mK
stage.

Special care was taken to avoid stray microwave radiation by using an outer and
an inner copper shield enclosing the device. The inner surface of
both shields was treated with commercially available Aeroglaze Z306 paint
\citeS{saeroglaze} absorbing far infrared stray radiation
\citeS{s:/content/aip/journal/rsi/70/5/10.1063/1.1149739}. We note that the
inner shield was not present for device N.

A typical DC $V-I$ trace of device S2 at $12\,$mK temperature is presented in
Fig.~S2b exhibiting a sharp transition between the dissipationless and the resistive
state at the switching current, $I_\textrm{sw}$. We observe a retrapping
current $I_r \ll I_\textrm{sw}$ characteristic to unshunted Josephson junctions
in the low temperature limit \citeS{sMassarotti2012}. The additional 
features in the resistive state are consistent with Fiske steps
\citeS{sPhysRev.138.A744}.

The switching current histograms were acquired using a Rigol DG4062 arbitrary
waveform generator controlling the isolated current bias source with a triangle
wave signal resulting in a $dI/dt$ current ramp.
A finite voltage response above the preset $V_\textrm{ref}\sim 10\,\mu$V
triggers the recording of $I_\textrm{sw}$ (Fig.~S2c). We note that the delay of
the low pass filters were accounted for on the basis of circuit simulations. Subsequent
$I_\textrm{sw}$ measurements were taken with setting zero $I_\textrm{bias}$ for
approximately $100\,$ms in between to avoid overheating effects. We did
not observe a difference in the switching current histograms taken with waiting
times in the range of $20\,$ms and $10\,$s.

\section{Basic characterization of the CPT}

In Fig.~S3a we present a typical charge stability diagram of the CPT.
First, we establish the superconducting gap of the aluminium lead and NbTiN
island by finding the onset of quasiparticle transport (Fig.~S3b and c).

\begin{figure}[ht!]
\centering
\includegraphics[width=\textwidth]{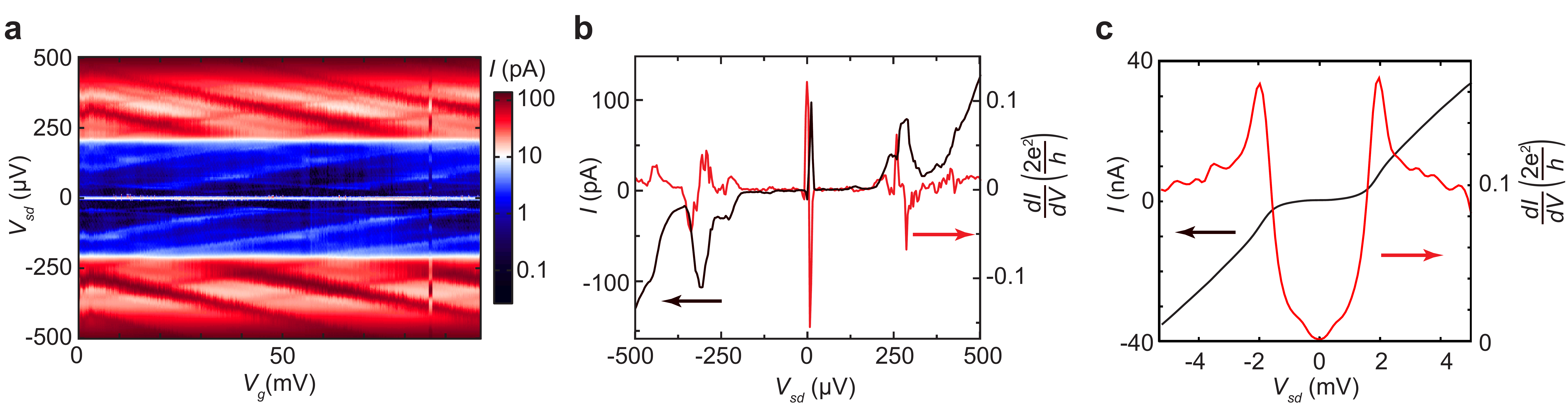}
\caption{\textbf{Basic characterization of the CPT.} \textbf{(a)} Stability
diagram: $\log|I|$ as a function of $V_\textrm{sd}$ and $V_g$. We estimate $E_c$
based on the characteristic resonances (see text). \textbf{(b),(c)} $I-V$
(black) and $dI/dV$ (red) line traces on the scale of $\Delta_\textrm{Al}$ and
$\Delta_\textrm{NbTiN}$, respectively. All data was taken for device S1.}
\end{figure}

\begin{table*}[!ht]
\centering
\begin{tabular}{| c || c | c | c | c | c | c | c |}
\hline
device & $\Delta_{Al}$ & $\Delta_{NbTiN}$ & $R_N$ &$E_J$ & $E_c$ &
$E_J/E_c$ & $E_J/E_c$\\
& ($\mu$eV) & ($\mu$eV) & (k$\Omega$) & ($\mu$eV) & ($\mu$eV) & calc. &
$I_{sw}$\\
\hline
\hline
N & 210 & 1390 & 58 & 54 & 50 & 1.08 & 1.25 \\
S1 & 218 & 1420 & 125 & 21 & 62 & 0.34 & 0.32 \\
S2 & 220 & 1300 & 66  & 48 & 49 & 0.98 & 1.16 \\
\hline
\end{tabular}
\caption{\textbf{Device transport parameters.}}
\end{table*}

It is important to note that we find a finite subgap conductance at
$eV_\textrm{sd} < 2\Delta_\textrm{NbTiN}$ which is consistent with the presence
of the subgap quasiparticle states justifying the analysis leading to
equation (1) in the main text. Furthermore, we observe reduced
$\Delta_\textrm{NbTiN}\approx1.3\ldots1.4\,$meV values compared to that of bulk
films ($\Delta>2\,$meV) \citeS{s1212327}, which we attribute to the chemical
interaction between the AlO$_x$ tunnel barrier and the NbTiN film. Indeed, it
was shown earlier that the critical temperature of Nb films is particularly very
sensitive to contamination with oxygen \citeS{sPhysRevB.9.888, sHalbritter1987}.
However, the nitridized NbTiN compound is presumably less prone to oxidization
\citeS{sDarlinski1987}.

We evaluate the Josephson coupling for a \emph{single} tunnel junction
\citeS{sPhysRevLett.10.486}:
\begin{equation}
E_J=\frac{\hbar}{2e^2}\frac{\Delta_\textrm{Al}}{R_N/2} \textrm{K}\left
(\sqrt{1-\left
(\frac{\Delta_\textrm{Al}}{\Delta_\textrm{NbTiN}}\right)^2}\right)
\end{equation}
with $K(x)$ being the complete elliptic integral of the first kind. This
expression is valid in the zero temperature limit, assuming symmetric tunnel
junctions of the resistance of \(R_N/2\).

We estimate the charging energy, $E_c=e^2/2C$, based on the periodicity of
characteristic resonances visible for $eV_\textrm{sd} \le 2\Delta_\textrm{Al}$
(Fig.~S3a) \citeS{sPhysRevLett.73.1541, sPhysRevB.76.172505}.

Alternatively, we can estimate the $E_J/E_c$ ratio based on the modulation of
$I_\textrm{sw}$ as a function of the gate charge, $n_g$ (last column of Table
S2) \citeS{sPhysRevLett.70.2940}. We find a good agreement between the $E_J/E_c$
values acquired by the two independent methods.

\section{Evaluation of the parity lifetime}

\begin{figure}[!ht]
\centering
\includegraphics[width=0.5\textwidth]{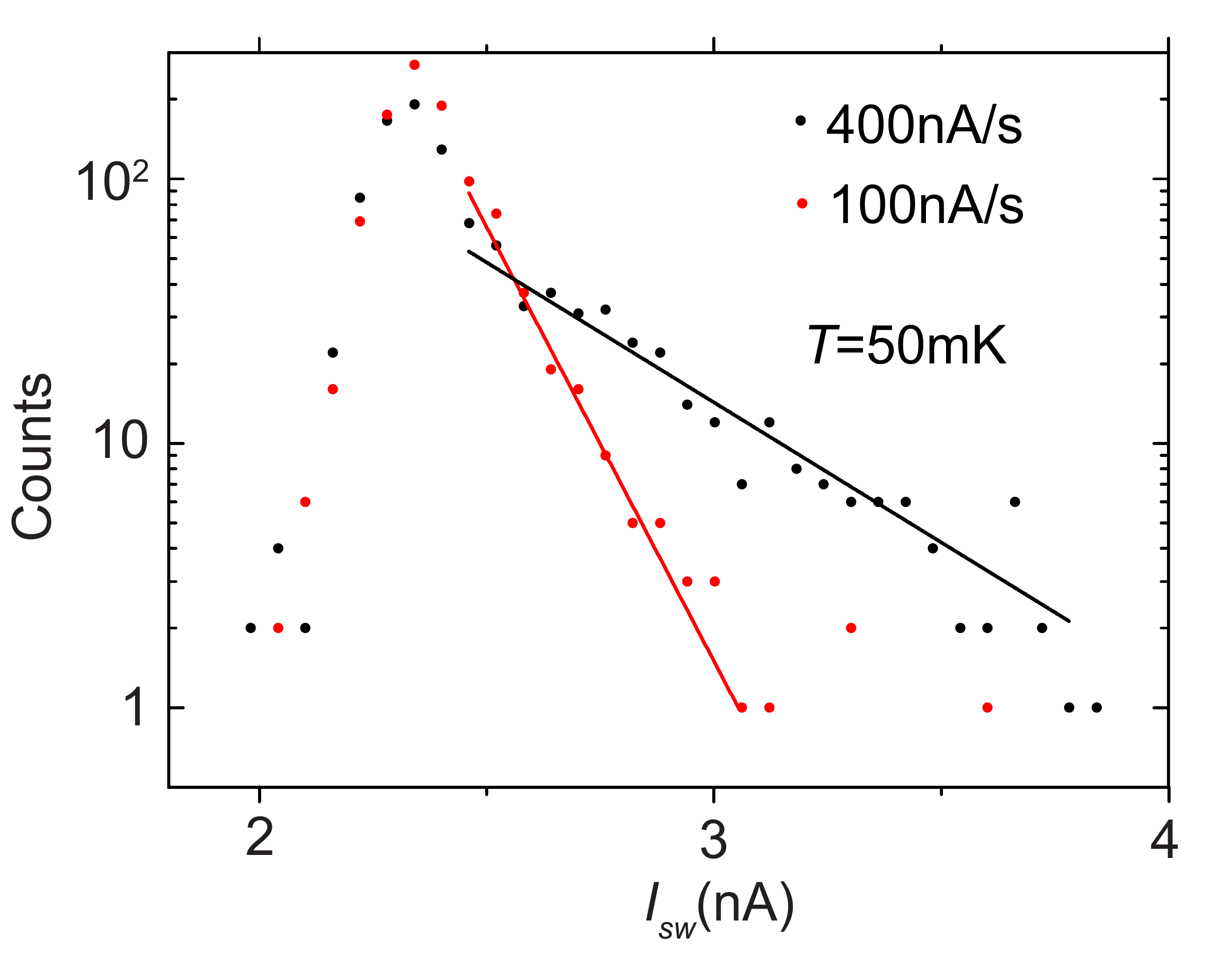}
\caption{\textbf{Measured switching current histograms at different $dI/dt$
ramp rates.} Colour legend: red: $dI/dt=100\,$nA/s, black: $dI/dt=400\,$nA/s.
The data was acquired on a CPT similar to device N.}
\end{figure}

We checked the robustness of the extracted parity lifetime against changing the
current ramp rate. Typical datasets are shown in Fig.~S4, giving \(\tau_p=1.2\,\)ms and
\(\tau_p=0.98\,\)ms for $dI/dt=100\,$nA/s (red) and $dI/dt=400\,$nA/s (black),
respectively. We estimate the typical uncertainty to be $15\%$, concluding
that $\tau_p$ does not depend on $dI/dt$ which validates the analysis in the main
text.
However, we do not discuss here the intrinsic peak shapes of the bimodal
switching current histogram. Since fast gate charge noise influences the
measured distribution \citeS{sjoyezthesis}, we cannot distinguish between
thermally activated \citeS{sPhysRevB.9.4760} and macroscopic quantum tunneling
behaviour \citeS{sClarke2008, sMassarotti2012}.

We now turn to the temperature dependence of $\tau_p$. In order to get
equation (1) in the main text, we assume the following:
\begin{enumerate}[i)]
  \item The superconducting gap of the island ($\Delta_\textrm{NbTiN} \gtrsim
  1.3\,$meV) is much higher than the effective thermal energy describing the
  quasiparticle population and the gap of the leads ($\Delta_\textrm{Al} \approx
  200\,\mu$eV).
  \item The density of states in the leads is BCS-type:
  $\rho_\textrm{lead}(E)=\rho_\textrm{Al} \times
  |E|/\sqrt{E^2-\Delta_\textrm{Al}^2}$ for $|E|>\Delta_\textrm{Al}$ and zero
  otherwise.
  \item There is a constant, finite subgap density of states
  $\rho_\textrm{subgap}$ for energies below $\Delta_\textrm{NbTiN}$ in the
  island.
  \item the energy dependence of the single electron tunnel probability is
  negligible over the energy range of $\sim \Delta_\textrm{NbTiN}$,
  meaning that the tunnel barrier is much higher than $\Delta_\textrm{NbTiN}$.
  \item The tunnel barriers are identical, each characterized by half the
  normal state resistance of the full device, $R_N$.
\end{enumerate}

Considering only single electron tunneling and zero voltage bias across the
tunnel barriers, following \citeS{sPhysRevB.85.012504}, we get the quasiparticle
tunnel rate:
\begin{equation}
\tau_p^{-1}= \frac{2}{e^2 (R_N/2) \rho_\textrm{Al}}
\frac{\rho_\textrm{subgap}}{\rho_\textrm{NbTiN}}\int_0 ^\infty
\rho_\textrm{lead}(E) f(T,E) dE =
\frac{2n_\textrm{qp}}{e^2R_N\rho_\textrm{Al}}\frac{\rho_\textrm{subgap}}{\rho_\textrm{NbTiN}.}
\end{equation}
 Assigning an effective temperature to the quasiparticle population in the
leads, we find:
\begin{equation}
n_\textrm{qp}(T)=2 \int_0 ^\infty \rho_\textrm{lead}(E) f(T,E) dE =
\rho_\textrm{Al}
\sqrt{2 \pi k_B T \Delta_\textrm{Al}}e^{-\frac{\Delta_\textrm{Al}}{k_B T}}
\end{equation}
in the limit of $k_B T \ll \Delta_\textrm{Al}$. For temperatures exceeding
$100\,$mK, we assume that the quasiparticle population is in thermal equilibrium, and
therefore the lattice temperature is equivalent to the effective quasiparticle
temperature: $\tau_p^{-1}(T) \propto \sqrt{T}\exp(-\Delta_\textrm{Al}/k_B T).$

We verify this picture by fitting the observed parity lifetimes with
$\Delta_\textrm{Al}$ as a free parameter, and find values ranging $170 \ldots
210\,\mu$eV for different devices in good correspondence with the gap determined
by voltage bias spectroscopy ($\Delta_\textrm{Al}$ in Table S2).

Notably, the ratio $\rho_\textrm{subgap}/\rho_\textrm{NbTiN}$ is the Dynes
parameter \citeS{sPhysRevLett.41.1509} of the island material, characterized to be
$\lesssim 10^{-3}$ based on measurements of highly resistive single junctions.
With this value and using $\rho_\textrm{Al}=1.45 \times
10^{47}\,\textrm{J}^{-1}\textrm{m}^{-3}$ \citeS{sPhysRevB.85.012504}, we get
$n_\textrm{qp} \sim 3\,\mu\textrm{m}^{-3}$ for device N based on the observed
parity lifetime of $9.5\,$ms in the low temperature limit.

We now comment on the observed $\Delta^\star \approx 20\,\mu$eV activation
energy observed for devices S1 and S2. We estimate the maximal even-odd energy
difference to be $\delta E \approx 20\ldots 30\,\mu$eV based on $E_J\approx
E_c\approx 50\,\mu$eV \citeS{sPhysRevB.75.212501} which is in range of the
experimentally observed $\Delta^\star$. Similar, activated behaviour of the
parity lifetime scaling as $\sim \exp(\delta E/k_B T)$ was reported earlier
\citeS{sPhysRevB.77.100501}. 

Providing another possible explanation, we note that a grain size of
$\approx 50\,$nm can lead to a level spacing of the order of $10\,\mu$eV which
can influence single electron transport and hence $\tau_p$ if the grains are
weakly coupled, i.~e.~for disordered superconducting films \citeS{sdubi2007}.
Disorder-induced fluctuations may also explain the broadening of the coherence
peaks (Fig.~S3c) \citeS{ssacepe2011, sPhysRevB.88.180505}.

\section{Superconducting thin film characterization and magnetic field dependence}

Next, we consider the properties of superconducting stripes with
layer thickness $d$, and a width $w$ to find the London penetration depth
$\lambda_L$ and the coherence length $\xi$. We characterize the upper critical
field in the parallel ($B_{c\parallel}$) and perpendicular geometry
($B_{c\perp}$) based on the $dI/dV$ traces of the tunnel junctions of the CPT.
In addition, we measure the normal state resistivity of the films that gives an
estimate for the mean free path, $l_m$ \citeS{skittelbook}.

First, we establish the length scales of the island material, NbTiN. We find
films superconducting at \(B_\perp=9\,\)T which leads to an
upper limit of \(\xi_\textrm{NbTiN}<6\,\)nm following \citeS{stinkhambook}:
\begin{equation}
B_{c\perp}=\frac{\Phi_0}{2\pi \xi^2}.
\end{equation}
The penetration depth can be estimated using the normal
state resistivity of \(\rho=98\,\mu\Omega\)cm and the critical temperature of
$T=14.1\,$K using the following semi-empirical formula \citeS{s1212327}:
\begin{equation}
\lambda_\textrm{NbTiN}=\sqrt{\frac{\rho [\mu\Omega\textrm{cm}]}{T_c [K]}}\times
105\,\textrm{nm} \approx 280\,\textrm{nm}.
\end{equation}

Next we estimate length scales of the Al leads based on the electronic transport
through the CPT. Typical thin Al films are type-II superconductors in the dirty
limit ($l_m < \xi_0$) with a reduced coherence length of
\(\xi\approx0.85\sqrt{\xi_0 l_m}\) and with a London penetration depth of
\(\lambda\approx\lambda_0 \sqrt{\xi_0/l_m}\), where \(\xi_0\approx1500\,\)nm and
\(\lambda_0\approx 16\,\)nm are the bulk values \citeS{stinkhambook}. For our
films, we estimate \(l_m\approx8.5\,\)nm based on the resistivity of
\(\rho=4.3\,\mu\Omega\)cm \citeS{skittelbook}. From the stability diagram of the
devices, we extract upper critical fields of $B_{c2,\perp}=36.4\pm4\,$mT
(Fig.~S5d) and $B_{c2,\parallel}=320\pm10\,$mT (Fig.~S5b) leading to a coherence
length of $\xi_\textrm{Al}=96\,$nm and $\lambda_\textrm{Al}=230\,$nm which
enables vortex formation in the aluminium leads in perpendicular magnetic field.

\begin{figure}[ht!]
\centering
\includegraphics[width=\textwidth]{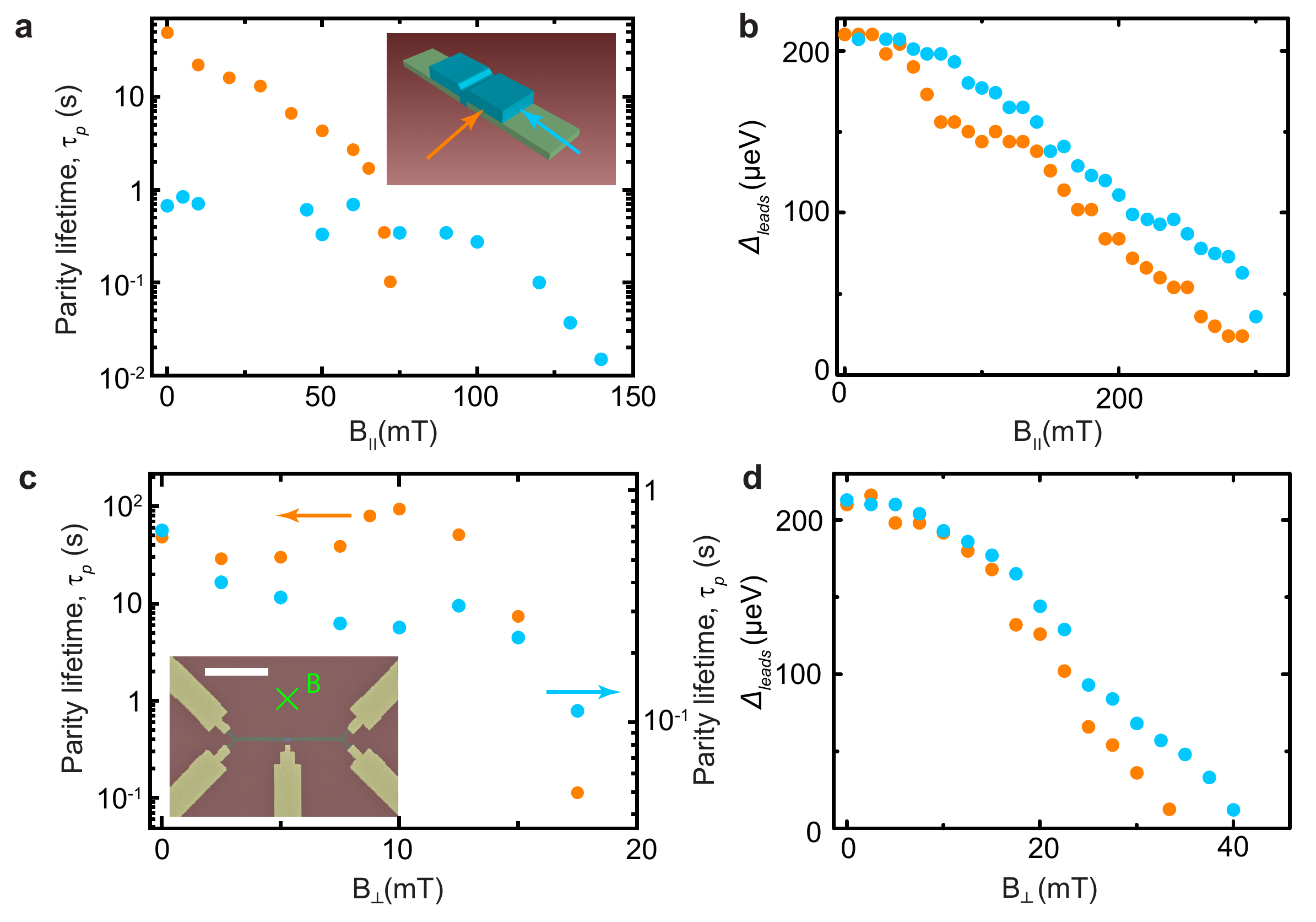}
\caption{\textbf{Parity lifetime and superconducting gap in magnetic
field.} \textbf{(a)} Measured parity lifetime as a function of in-plane magnetic
field. The inset shows the relative orientation of the field for the two
devices. \textbf{(b)} The superconducting gap of the Al leads as a function of
in-plane magnetic field. \textbf{(c),(d)} The parity lifetime and the
superconducting gap as a function of the perpendicular magnetic field. Colour
legend: cyan: device S1; orange: device S2. The scale bar in the inset of
panel (c) denotes $5\,\mu$m.}
\end{figure}

It is important to observe that $\xi_\textrm{NbTiN} \ll d\approx 100\,$nm
enables vortex formation for an \emph{in-plane} geometry in the NbTiN island. We
find a characteristic suppression of $\tau_p$ at \(B_\parallel=70\,\)mT for
device S2 (orange dots in Fig.~S5a) and at \(B_\parallel=110\,\)mT for device S1 (cyan
dots in Fig.~S5a). Considering the effective cross-sectional areas (see Table S1
for dimensions), we find \(\Phi\approx 1.5 \Phi_0\) and \(\Phi\approx 1.1
\Phi_0\) for S2 and S1, respectively, which is in qualitative agreement with the
threshold of a single vortex formation in a mesoscopic island
\citeS{sMoshchalkov1995, sGeim1997, sPhysRevB.58.R5948}. We also
note that $B_{c2,\parallel}$ of the leads (Fig.~S5b) does not depend on the
direction of $B_\parallel$, therefore the different evolution of $\tau_p$ can
only be explained by the different alignment of $B_\parallel$ with respect to
the NbTiN islands.

In a perpendicular geometry, the vortex phase is stable in a thin stripe above
the magnetic field
\begin{equation}
B_{c1,\perp}(w)=\alpha \frac{\Phi_0}{w^2},
\end{equation}
where $\alpha$ is a model-dependent prefactor \citeS{sPhysRevLett.92.097003,
sPhysRevB.84.220502, sPhysRevB.51.3092} of the order of unity. We reproducibly
find the same non-monotonic behaviour of $\tau_p$ for devices S1 and S2 with the maximum at
$B_\perp \approx 10\ldots13\,$mT, which is in range of $B_{c1,\perp}(w)$ for
$w=200\ldots250\,$nm, the width of the Al leads near the island.

\bibliographyS{references_supplement}

\end{document}